# Improved weather generator algorithm for multisite simulation of precipitation and temperature

Leanna King[1], A. Ian McLeod[2] and Slobodan P. Simonovic[1]

[1] Department of Civil and Environmental Engineering, The University of Western Ontario, London, Ontario, Canada, N6A 5B9, (519) 872-6917, leannamichelleking@gmail.com

[2] Department of Statistical and Actuarial Sciences, The University of Western Ontario, London, Ontario, Canada

**Abstract**

The KnnCAD Version 4 weather generator algorithm for nonparametric, multisite simulations of temperature and precipitation data is presented. The K-nearest neighbour weather generator essentially reshuffles the historical data, with replacement. In KnnCAD Version 4, a block resampling scheme is introduced to preserve the temporal correlation structure in temperature data. Perturbation of the reshuffled variable data is also added to enhance the generation of extreme values. A case study of the Upper Thames River Basin in Ontario, Canada is performed and the model is shown to simulate effectively the historical characteristics at the site. The KnnCAD Version 4 approach offers a major advantage over parametric and semi-parametric weather generators as it can be applied to multiple sites for simulation of temperatures and precipitation amounts without making assumptions regarding the spatial correlations and the probability distributions for each variable.

**Introduction**

Weather generators are stochastic simulation tools that are commonly used to produce synthetic climate data of any length with the same characteristics as the input record. Such algorithms are often used for hydrological applications [Dibike and Coulibaly, 2005; Charles et al., 2007; Kwon et al., 2011]. In recent years, their application has been extended for statistical downscaling of Atmosphere-Ocean coupled Global Circulation Model (AOGCM) outputs to investigate the impacts of climate change at a basin scale [Dibike and Coulibaly, 2005; Hashmi et al., 2011; Eum and Simonovic, 2011].

There are several categories of stochastic weather generators. Parametric models typically follow the WGEN approach [Richardson, 1981; Craigmile and Guttorp, 2011; Soltani and Hoogenboom, 2003; Kuchar, 2004], which uses Markov chains to simulate wet and dry spells and probability distributions for temperature and precipitation amounts. Some other parametric weather generator examples include SIMMETEO [Geng et al., 1988; Soltani and Hoogenboom, 2003; Elshamy et al., 2006], WGENK [Kuchar, 2004], AAFC-WG [Qian et al., 2004; 2008] and GEM [Hanson and Johnson, 1998]. A drawback to the parametric models is that they require careful statistical checks to ensure the developed probability distributions are suitable to the study area [Sharif and Burn, 2006]. Furthermore when the models are used with low-order Markov dependence, they cannot adequately simulate wet and dry spell lengths, underestimating the occurrence of prolonged drought or rainfall events [Semenov and Barrow, 1997; Dibike and Coulibaly, 2005; Sharif and Burn, 2007; Mehrotra and Sharma, 2007].

As a result of the parametric model limitations, there are several semi-parametric models. Two of the commonly used models are SDSM [Wilby and Dawson, 2007] and LARS-WG [Semenov and Barrow, 2002]. A drawback to both of these models is that they may only be used for one site at a time and spatial correlations must be assumed for multisite applications. SDSM is a regression-based model where linear regression relationships are developed between large scale atmospheric predictor variables from reanalysis datasets and locally observed station data. SDSM has been successfully applied to simulate temperature and precipitation data in many regions around the world [Khan et al., 2006; Chen et al., 2010; Hashmi et al., 2011; Liu et al., 2011].

LARS-WG is a semi-empirical model that uses the historical station record to develop probability distributions for wet and dry spells and semi-empirical distributions for temperature and precipitation amount conditional on the spell lengths. Some studies have found that LARS-WG underestimates the occurrence of extreme temperature events [Mavromatis and Hansen, 2001; Qian et al., 2004; Semenov, 2008], However, LARS-WG has been successfully applied for climate data simulation in various study areas [Khan et al., 2006; Rivarola Sosa et al., 2011; Abbasi et al., 2011].

Due to the limitations associated with parametric and semi-parametric weather generators, nonparametric models have become increasingly popular. Nonparametric weather generators can create multisite, multivariate climate simulations without making assumptions regarding the inter-site spatial correlations or the probability distributions of the variables [Sharif and Burn, 2006; 2007]. Nonparametric models typically use a K-Nearest Neighbour (K-NN) procedure which resamples from the historical record, with replacement [Yates, 2003; Sharif and Burn, 2006; Eum et al., 2010]. The nearest neighbours to the current day are selected by calculating the Mahalanobis distance metric between the current day and each of the days within a temporal window centered on that day from the entire $N$ years of record. The closest $K$ days are retained and one is randomly chosen as the next day's weather, with a higher probability given to closer days [Yates, 2003; Sharif and Burn 2006; 2007].

One limitation to the K-Nearest Neighbour approach is that all predictor variables are considered equally important during the selection of the $K$ closest neighbours. This may not be the case in reality, where certain variables are more important than others and some can even be inter-correlated and therefore less significant, resulting in a biased list of the $K$ closest neighbours [Mehrotra and Sharma, 2006]. This is also related to the "curse of dimensionality", when the number of dimensions (variables and stations) for K-NN simulations increase, the concept of a nearest neighbor can become meaningless [Hastie et al., 2008]. In order to minimize this bias, the current work employs only three variables: Precipitation amount, maximum temperature and minimum temperature. Furthermore, a regional average is used to minimize the dimensions when choosing a nearest neighbor [Eum and Simonovic, 2011].

Traditional K-NN approaches have been found to underestimate the occurrence of wet and dry spells [Apipattanavis et al., 2007]. Another drawback is that the reshuffling procedure results in a loss of the temporal correlation structure of daily temperatures. The extreme values in the output data are also limited by those in the input dataset since the values are resampled [Yates, 2003]. Sharif and Burn [2006; 2007] developed a perturbation method to generate alternative extreme values for precipitation in KnnCAD Version 1, however Eum and Simonovic [2011] found that their methodology could not be easily extended for generation of alternative temperature extremes. Lee et al., [2012], used a gamma kernel density estimate to successfully perturb the reshuffled precipitation data; however their approach could not easily be used for temperature outputs.

In this study, the KnnCAD Version 4 algorithm is presented to address the issues associated with the previous versions of the algorithm. A block resampling procedure is added to improve temporal correlations of temperature variables. A perturbation procedure is also introduced to enhance the generation of extreme temperature and precipitation values. The model outputs for the CLIMDEX set of statistics [CLIMDEX, 2012] are presented in a comparison with weather generators LARS-WG and SDSM. The following section presents the details of the KnnCAD Version 4 weather generator algorithm. Next is an application of the weather generators to the Upper Thames River basin in Ontario, Canada. Finally conclusions of the paper are presented.

**Methods**

The KnnCAD Version 4 weather generator is an extension of the model of Yates [2003]. The model essentially reshuffles the observed daily data, so application to multiple sites is achieved by selecting the corresponding day's weather at all stations. In this way, the spatial correlations of the climate variables are inherently preserved.

The model proceeds in steps by creating a subset of days from each year in the historical record that are centred within a temporal window on the current day of the simulation. The current day is removed to prevent repeated daily values. This subset of "potential neighbours" has length $L=N*(w+1)-1$ for $N$ years of record and a temporal window of length $w$. The regional average from all stations is computed for each variable and day in the potential neighbours. These potential neighbour averages are then compared to the current day's regional average using a distance metric; the Mahalanobis distance [Yates, 2003; Sharif and Burn, 2006]. Based on their distance from the current day, the potential neighbours are ranked and the first $K$ are selected, the "K-nearest neighbours". Based on the days' ranks, a cumulative probability distribution is developed. The next day's weather is then selected by generating a random number $u(0,1)$ and comparing this to the probability distribution, selecting the closest day. As such, days which are more similar to the current day have a greater probability of selection. $B$ days following the resampled day in the historical record are also resampled in order to preserve the temporal correlation structures of variables such as temperature. Each of the resampled values is then perturbed to ensure unique values are generated that do not necessarily occur in the historical record. The simulation proceeds until a dataset of reshuffled values is generated, with the same length as the input data set. The simulation can be repeated several times to generate ensembles of synthetic daily data. AOGCM simulations can be generated by applying monthly change factors to the input dataset. Detailed steps of KnnCAD V4 are presented below.

(1) Compute the regional means of *p* variables (*x*) across all *q* stations for each day (*t*) in the historic record of length *T*, following Equations (1) and (2):

$$\overline{X}_t = \left[\overline{x}_{1,t}, \overline{x}_{2,t}, \ldots, \overline{x}_{p,t}\right] \quad \forall t = \{1, 2, \ldots, T\} \tag{1}$$

$$\text{Where } \overline{x}_{i,t} = \frac{1}{q}\sum_{j=1}^{q} x_{i,t}^{j} \quad \forall i = \{1, 2, \ldots, p\} \tag{2}$$

The variables that are typically used in the KnnCAD V4 approach are precipitation, maximum temperature and minimum temperature. Other variables may also be considered in some applications.

(2) Choose a temporal window of length *w*, and select a subset of *L* potential neighbours to the current day of simulation, where $L = N*(w+1)-1$ for *N* years of record. The potential neighbours are the days within the temporal window centered on the current day of the simulation, *t*, and contain *p* variables for a total of *L* days,. Yates [2003] used a temporal window of 14 days in the Great Lakes region, so if January 20[th] is the current day, the potential neighbours are all days that fall between January 13[th] and January 27[th] for all *N* years, excluding the value of the current day to prevent repeating weather sequences.

(3) Compute the regional means $\overline{X}_l$, of the *L* potential neighbours (*l*=1,2,…,*L*) for each day across all *q* stations.

(4) Compute the covariance matrix, $C_t$, for day $t$ using the potential neighbours from (3) with a standardized data block of size $L$ by $p$.

(5) Randomly choose $p$ variables at $q$ stations from one of the $N$ years of historical record for the first time step day (eg. January 1).

(6a) Calculate the eigenvector and eigenvalue from the covariance matrix $C_t$.

(6b) Retain the eigenvector $E$ which corresponds to the highest eigenvalue which explains the largest fraction of variance in the p variables.

(6c) Calculate the first principal component using $E$ from (6b):

$$PC_t = \bar{X}_t E \qquad (3)$$

$$PC_l = \bar{X}_l E, \quad \forall l = \{1, 2, \ldots, L\} \qquad (4)$$

Where $PC_t$ and $PC_l$ are one-dimensional values transferred from the eigenvector in (6b) for the current day, $t$ and the $l^{th}$ of $L$ potential neighbors. Only one principal component is retained following the recommendation of Eum and Simonovic [2008].

(6d) Calculation of the Mahalanobis distance using the values obtained in Equations (3) and (4) as well as the variance, Var($PC$), between all L values of $PC_l$.

$$d_l = \sqrt{\frac{(PC_t - PC_l)^2}{Var(\boldsymbol{PC})}} \qquad \forall l = \{1, 2, \ldots, L\} \qquad (5)$$

(7) Select the number $K$ of nearest neighbors to retain out of the $L$ potential values. Rajagopalan and Lall [1999] and Yates et al. [2003] recommend taking $K = \sqrt{L}$. Sort the Mahalanobis distance metric from smallest to largest, and retain the first $K$ neighbours on the list. Use a discrete probability distribution weighting closest neighbours highest for resampling one of the $K$ values.

(8) Generate a random number, $u(0,1)$ and compare this to the cumulative probability, $p_m$, to determine the current day's nearest neighbor. The day $m$ for which $u$ is closest to $p_m$ is selected as the nearest neighbor and the corresponding weather is used for all stations in the region. Through this step, spatial correlation among the variables is preserved.

(9) Resample $B$ days from the historical record which follow the selected day ($m$) from step (8). $B$ is the block length and is selected based on the observed autocorrelations of daily temperatures (lag 1, lag 2, etc.). It should be as large as required for the model to reproduce the observed temperature autocorrelations

(10) Perturbation of the reshuffled variable values for days $t$ to $t+b$ (where $b=1,2,\ldots,B$). A new perturbation component is introduced to ensure simulation of unique but reasonable values for temperature and precipitation amount that can lie outside of the observed ranges. Because precipitation has a non-negativity constraint, it must be dealt with differently from temperature.

As such, the same interpolation equation is used for both precipitation and temperature with different randomly distributed variables, $Z_{t+b}$, as shown in steps (10a) and (10b).

(10a) Perturbation of the reshuffled temperature values $x^j_{i,t+b}$ for temperature variable $i$, station $j$ and day $t+b$ following Equation (6):

$$y^j_{i,t+b} = \lambda_{temp} x^j_{i,t+b} + (1 - \lambda_{temp}) Z_{t+b} \tag{6}$$

Where $y^j_{i,t+b}$ is the simulated perturbed value and $\lambda_{temp}$ is chosen between 0 and 1 during calibration (1 gives an unperturbed result and 0 yields a result based entirely on perturbation). For preservation of temporal correlations, $\lambda_{temp}$ should be as large as is reasonable. $Z_{t+b}$ is a normally distributed value with a mean of $x^j_{i,t+b}$ and a standard deviation of $\sigma^j_{i,t}$, calculated from the K-nearest neighbours for day $t$, station $j$, and temperature variable $i$. To prevent minimum temperature from exceeding maximum temperature, the same random normal variable $z$ is used for both maximum and minimum temperature across all stations and its value is transformed using the variables' corresponding $x^j_{i,t+b}$ and $\sigma^j_{i,t}$ values.

(10b) Perturbation of the reshuffled non-zero precipitation values $x^j_{ppt,t+b}$ for station $j$ and day $t+b$ (where $b=1,2,\ldots,B$), following Equation (7):

$$y_{ppt,t+b}^{j} = \lambda_{ppt} x_{ppt,t+b}^{j} + (1 - \lambda_{ppt}) Z_{t+b} \tag{7}$$

Where $Z_{t+b}$ comes from a two-parameter lognormal distribution. The parameters are calculated using the method of moments from Singh [1998]. The mean is equal to the unperturbed precipitation value and the standard deviation equals that of the non-zero values in the potential neighbors subset (the days that lie within a temporal window centered on the current day *t*). $\lambda_{ppt}$ is chosen between 0 and 1 and should be as large as is reasonable to preserve spatial correlation. The proposed perturbation scheme inherently produces values above zero while still producing perturbed precipitation amounts that can be either higher or lower than the unperturbed value.

(11) Repeat steps 6 through 10 until the end of the historical record is reached. Multiple simulations can be done to produce long synthetic datasets.

The KnnCAD program is coded in R programming language and has a Visual Basic decision support system to aid researchers in applying the program for their study area. The user may vary parameters such as the block length, *B*, and the interpolation parameters, $\lambda_{temp}$ and $\lambda_{ppt}$, to determine which combination of parameters provides the best calibration based on the outputs provided by the decision support system. To generate future climate scenarios from AOGCM models, the user can apply monthly change factors to the historical daily station data.

**Application**

To illustrate the utility of the KnnCAD V4 model, a case study of the Upper Thames River Basin (UTRB) is presented. KnnCAD V4 outputs are compared with outputs from SDSM and LARS-WG to demonstrate its utility as a weather generator. The UTRB, shown in Figure 1, is located between the Great Lakes of Erie and Huron and has a population of 515,640, the majority residing in London, Ontario, the major urban center in the region. The basin has a major history of flooding events, often occurring in March or April following snowmelt. Floods can also occur after a sudden peak in temperatures during the winter or in the summer after extreme precipitation events [Wilcox, 1998]. There have been a number of studies assessing the potential impacts of climate change on the UTRB, indicating vulnerability of the basin to future extreme precipitation events and flooding [Sharif and Burn, 2006; 2007; Solaiman et al., 2010; Simonovic, 2010; Eum and Simonovic, 2011; King et al., 2012].

Figure 1: The Upper Thames River Basin

Table 1: Upper Thames River Stations

A total of 22 stations around the basin are used in this study based on the availability and completeness of the Environment Canada datasets. A 27-year historical record from 1979-2005 is gathered from each station. The selected record length is chosen based on the availability and completeness of Environment Canada data at each of the 22 weather stations. The station locations can be found in Figure 1. Table 1 provides the names, elevations, latitudes and longitudes of each station. The historical data from each of these stations is used as input to the KnnCAD Version 4 algorithm which is then used to produce 25 ensembles of synthetic historical climate data. Each ensemble has the same length as the input dataset (27 years) so in total 675 years of synthetic historical climate data is output by the weather generator. It is important to generate a number of ensembles due to the random component of the program as each simulation is different. By varying the parameters for block length and interpolation, it was found that $B=10$ days and $\lambda_{ppt} = \lambda_{temp} = 0.9$ provided the best calibration result. For more details on the validation procedure for KnnCAD V4, please refer to King [2012].

The ability of the algorithm to simulate historical climate characteristics is investigated in terms of total monthly precipitation, daily precipitation characteristics, extreme precipitation events, wet spell lengths as well as mean and extreme daily temperatures. Root mean square errors for each of these indices are computed. The Wilcoxon test for equality of means and the Levene's test for equality of variances is also performed between the daily historical and simulated data for each month.

In order to demonstrate the utility of KnnCAD V4 as a climate simulation tool, the results for the London A station have been compared with simulated outputs from SDSM and LARS-WG, which are two widely used weather generators. For LARS-WG, precipitation, solar radiation and maximum and minimum temperatures are used as inputs and the model is easily calibrated for the London A station to produce a 675 year simulated output. For SDSM, several North American Regional Reanalysis predictor variables were screened, and it was found that mean sea level pressure and northward wind speed were the best predictors for precipitation. For temperatures, northward wind speed and specific humidity were used as predictors. Both SDSM and LARS-WG were validated using the first half of the data and comparing outputs to the second half of data. For a complete description of the calibration and validation of SDSM and LARS-WG at the London A station, see King [2012]; King et al., [2012]. The outputs from the fully simulated data are compared to KnnCAD V4 using root mean square errors and CLIMDEX indices.

Figure 2 shows the KnnCAD V4 precipitation results for total monthly precipitation and daily precipitation characteristics from the London A station in the UTRB. Results at other stations are similar. The figure on the left presents boxplots of simulated total monthly precipitation amounts, with the historical medians shown as a line plot. The upper and lower lines in the box represent the quartiles, the middle line represents the median and the whiskers extend to 1.5 times the interquartile range. It is clear from the figure that while there are slight over and underestimations in some months, KnnCAD Version 4 simulates values that are very close to the observed medians. The number of outliers indicates the ability of the model to generate extreme precipitation events. Results from the KnnCAD Version 3 of Eum and Simonovic [2011] are very similar and thus are not included for discussion. Please refer to King [2012] for this and other results from SDSM and LARS-WG simulations.

The graph on the right in Figure 2 is a dot plot showing the KnnCAD V4-simulated monthly values of daily precipitation standard deviations in dark blue and means in light blue. The observed values are shown as a line plot. Each dot in the figure represents the result from one ensemble of climate data, for a total of 25 dots in each grouping. Overall, the spread of points is generally centered on the historical observation indicating good performance of the model in daily precipitation simulation. There is a slight overestimation in the standard deviations for April and October but the mean values are simulated well. These results indicate the ability of the model to simulate effectively the characteristics of daily precipitation in the region. The results from KnnCAD Version 3 are similar.

Figure 2: Simulated and observed total monthly precipitation and daily precipitation characteristics at London A

Figure 3 shows the KnnCAD V4-simulated extreme daily precipitation events at London A on the left and wet spell lengths on the right, as dot plots with the historical observed data as line plots. In the figure on the left, the 99$^{th}$ percentile events are shown in dark blue and the 95$^{th}$ percentile events in light blue. There is a slight overestimation in the April results for the extreme percentile events and most months for the 99$^{th}$ percentiles contain an uncharacteristic point; however the spread of the points are generally centered on the historical observation. The results from KnnCAD V3 of Eum and Simonovic [2011] are very similar and thus are not included for discussion.

For the wet spell lengths (right), the dark blue points represent the KnnCAD V4-simulated monthly maximum wet spell lengths and the light blue points represent the mean wet spell lengths. It is clear that KnnCAD Version 4 performs very well as the points are centered on the historical observation in all months but the maximum wet spell length for October. Results for dry spell lengths are similar. This is likely due to the block resampling addition in the program which helps preserve wet and dry spell structure.

Figure 3: Simulated and observed extreme daily precipitation and wet spell lengths at London A

Figure 4 shows the results for extreme temperature values. Because the median values are simulated quite well by KnnCAD Version 4 they are not presented. In Figure 4, the dot plot on the right shows the daily extreme high temperatures (95$^{th}$ and 99$^{th}$ percentiles) of maximum temperatures in red and minimum temperatures in blue. The dot plot on the left shows extreme low temperatures (5$^{th}$ and 1$^{st}$ percentiles) of maximum and minimum temperatures in red and blue, respectively. It is clear in the figures that the simulated dots are very close to the observed value which is shown as a line plot. Results for KnnCAD V3 are similar, but it should be noted that Version 3 of the algorithm is unable to perturb the resampled temperature values and thus cannot produce alternative extreme values in the simulated temperatures.

Figure 4: Simulated and observed extreme daily maximum and minimum temperatures at London A

Table 2 shows a comparison of root mean square error (RMSE) values for selected indices from SDSM, KnnCAD V4 and LARS-WG. RMSE values are presented in the same units as the selected index. The table shows that KnnCAD V4 performance is better than both LARS-WG and SDSM for 14 out of the 19 indices presented. LARS-WG very slightly outperforms KnnCAD V4 for total monthly precipitation, with a difference in the RMSE of about 1mm. For the standard deviation of precipitation, LARS-WG scores an RMSE value of 0.08 less than that of KnnCAD V4, however both results indicate good performance. For the 95$^{th}$ percentile of precipitation, the differences are very small between LARS-WG and KnnCAD V4. For the 99$^{th}$ percentile, KnnCAD V4 is outperformed by about 0.76mm, however the RMSE values are both quite small indicating good performance of the models. SDSM consistently ranks 2$^{nd}$ or 3$^{rd}$ in terms of the RMSE values. For wet and dry spell lengths, KnnCAD V4 consistently outperforms the other two weather generators. This is likely a result of the block resampling addition to the algorithm which improves simulation of wet and dry spells. For all temperature indices except the mean daily temperatures, KnnCAD V4 outperforms the other models. Overall, the RMSE values indicate satisfactory performance of the KnnCAD V4 algorithm in simulation of selected indices.

Table 2: Root mean square error values for selected climate indices from three weather generators at London A station

    Figure 5 shows the lag-1 autocorrelations for maximum and minimum temperatures at the London A station. The column on the left presents the results for KnnCAD Version 4 which resamples 10 days at a time from the historical record. The column on the right shows results from KnnCAD Version 3, which resamples one day at a time. Maximum temperatures are shown in the top row and minimum temperatures in the bottom row. Observed values are shown as a line plot. It is clear that by adding block resampling, the ability of the model to preserve the temporal correlation structure in the observed record is significantly improved. This is an important factor in the UTRB where snow accumulation and melt lead to major flooding events; it is crucial that hydrologic models in the study basin have a temporally correlated series of temperature data to effectively simulate such events. Other weather generators, such as SDSM and LARS-WG, are inherently unable to simulate temporally correlated temperature series, due to their stochastic nature [King, 2012].

Figure 5: Simulated and observed lag-1 autocorrelations in maximum and minimum temperatures at London A from KnnCAD Versions 3 and 4.

Figure 6 shows the spatial correlations of maximum temperatures for each of the stations when paired with London A. The results are presented for January but results in other months and for other station pairs and variables are similar. It is clear from the figure that the KnnCAD Version 4 can effectively simulate the inter-site spatial correlations. While there is a slight underestimation in the correlations with Exeter, the dots are centered on the historical observations for the rest of the stations. This is an inherent advantage of KnnCAD over the parametric and semi-parametric weather generators; it is able to accurately simulate spatial correlations without making any statistical assumptions.

Figure 6: Spatial correlations for station pairs with London A for January.

CLIMDEX extreme climate indices [CLIMDEX, 2012] have been included in Table 3 to provide a comparison between KnnCAD, SDSM and LARS-WG. The mean values of selected indices are presented for the historical dataset, as well as for the simulated outputs from each of the three weather generators. The table provides a description of each index for clarity. In the index descriptions, PPT represents precipitation amount (mm), TMAX represents maximum temperature (˚C) and TMIN represents minimum temperature (˚C). For some of the monthly indices, mean annual values are shown in order to effectively compare outputs from the three weather generators.

For annual precipitation, the Table 3 results show that the KnnCAD V4 weather generator is outperformed by the other models, as it underestimates annual precipitation by about 35mm whereas SDSM and LARS-WG underestimate it by about 4mm and 6mm, respectively. KnnCAD V4 outperforms other models for the simulation of maximum dry spell length, and all models perform quite well simulating the maximum wet spell length. For the annual count of days with precipitation greater than threshold levels, KnnCAD V4 outperforms SDSM and LARS-WG at the higher thresholds (20mm and 25mm), but the other models do slightly better at simulating the lower threshold 10mm events. KnnCAD V4 performs significantly better than SDSM and LARS-WG in the simulation of annual total precipitation occurring on extreme precipitation days (where precipitation is greater than the 95$^{th}$ and 99$^{th}$ percentile events). The KnnCAD V4 model is able to simulate this extreme precipitation index within 2-4mm of the historical value, whereas LARS-WG tends to overestimate the amounts by up to 20mm, and SDSM underestimates the values by up to 40mm. All weather generators perform well in the simulation of the simple precipitation intensity index, but KnnCAD V4 slightly outperforms SDSM and LARS-WG. Overall, KnnCAD V4 performs very well for most precipitation indices and results are comparable or better than SDSM and LARS-WG in most cases.

  For temperature indices, all models slightly underestimate average temperatures by about 1.5˚C to 2˚C. KnnCAD V4 outperforms the other models slightly for simulation of monthly mean minimum temperature. For the daily temperature range, both KnnCAD V4 and LARS-WG perform very well, estimating the mean range within 0.2˚C of the historical value. For frost days, summer days and tropical nights, KnnCAD V4 outperforms LARS-WG and SDSM; however all models overestimate the number of frost days and underestimate the number of summer days and tropical nights. All models also overestimate the number of icing days by at least 10 days, with SDSM slightly outperforming the other models. All models underestimate the warm spell duration index and overestimate the cold spell duration index. Overall, the models could not accurately simulate indices such as the annual temperature threshold days and the hot/cold spell durations well. However, the KnnCAD V4 performance is comparable to SDSM and LARS-WG for all temperature indices.

Table 3: Historical and simulated mean values of CLIMDEX climate extreme indices from SDSM, KnnCAD V4 and LARS-WG.

Figure 7 shows the monthly maximum 1-day precipitation amounts from the historical record as well as results from KnnCAD V4, LARS-WG and SDSM. Each of the weather generators simulates the 1-day precipitation amounts quite well, with some slight over or underestimations, depending on the month. From November to April, all models underestimate the amount by 1mm to 7mm. Overall LARS-WG and KnnCAD V4 simulate these months better than SDSM, which underestimates the amounts more significantly. Results are more variable from May to October, with KnnCAD and LARS-WG overestimating the summer maximum 1-day precipitation amounts by up to 4mm and SDSM simulating these months quite well. All models are able to capture the general trend in the precipitation amounts fairly well with some slight over and underestimations.

Figure 7: Monthly maximum 1-day precipitation amounts from the observed data and the simulated KnnCAD V4, LARS-WG and SDSM results.

Figure 8 shows the maximum consecutive 5-day precipitation amounts (in mm) from the historical record as well as the simulated results from KnnCAD V4, LARS-WG and SDSM. Each of the weather generators simulates the monthly trends fairly well. There is generally less of an agreement between simulated and historical for the period from December to April, and more of an agreement in the summer months. Performance of KnnCAD V4 in the simulation of consecutive 5-day precipitation totals is comparable to LARS-WG and SDSM.

Figure 8: Monthly maximum 5-day consecutive precipitation amounts from the observed data and the simulated KnnCAD V4, LARS-WG and SDSM results.

Figure 9 shows the results for monthly daily temperature ranges from the three weather generators, plotted with the historical average in red. The historical trend is highly seasonal, with a greater daily temperature range in the summer months than in the winter. All models simulate the trend quite well, with LARS-WG and KnnCAD V4 performing slightly better than SDSM.

Figure 9: Monthly daily temperature range from the observed data and the simulated KnnCAD V4, LARS-WG and SDSM results.

**Conclusions**

The KnnCAD Version 4 model provides an improvement over previous versions of the weather generator by adding block resampling to improve the temporal correlation structure and perturbation to enhance the simulation of extremes. It is able to simulate effectively the historical climate variables at several sites simultaneously, without making statistical assumptions regarding variables' probability distributions and spatial correlations between weather stations. As such, KnnCAD Version 4 provides a major advantage over the semi-parametric and parametric weather generators. In a comparison with SDSM and LARS-WG models, it was found that KnnCAD Version 4 performance was comparable and in some cases better for selected climate indicators as well as CLIMDEX extreme weather indices at both the annual and monthly scale. One weakness of the current KnnCAD Version 4 model is that it can only be used on daily precipitation and temperature data, whereas SDSM and LARS-WG are able to simulate a larger variety of climate variables. An advantage KnnCAD V4 has over the other stochastic models is that it is able to reproduce the temporal correlation structures of the daily temperature series. This is of particular importance in study basins such as the Upper Thames River Basin, where snow accumulation and melt lead to major flooding events.

Currently, the KnnCAD Version 4 model is being validated for other climatic regions such as Brazil and in other areas of Canada. A new methodology for developing AOGCM-modified input datasets for KnnCAD Version 4 from daily AOGCM data is an important area for future research, as the current weather generator cannot create future climate scenarios that take into account changes in the variability of future daily temperatures and precipitation amounts.

**Acknowledgements**

Dr. Ian McLeod and Dr. Slobodan Simonovic acknowledge with thanks support from the NSERC Discovery Grant Program. The authors wish to thank Environment Canada for the climate data used in this study, as well as Ontario Graduate Scholarships (OGS) and the Canadian Federation for Climate and Atmospheric Sciences (CFCAS) for funding this research. The authors would also like to acknowledge Ms. Sarah Irwin for calculating CLIMDEX indices for the weather generator outputs.

**Tables**

Table 1: Upper Thames River Stations

| Station | Latitude (deg N) | Longitude (deg W) | Elevation (m) |
|---|---|---|---|
| Blyth | 43.72 | 81.38 | 350.5 |
| Brantford | 43.13 | 80.23 | 196.0 |
| Chatham | 42.38 | 82.2 | 198.0 |
| Delhi CS | 42.87 | 80.55 | 255.1 |
| Dorchester | 43.00 | 81.03 | 271.3 |
| Embro | 43.25 | 80.93 | 358.1 |
| Exeter | 43.35 | 81.50 | 262.1 |
| Fergus | 43.73 | 80.33 | 410.0 |
| Foldens | 43.02 | 80.78 | 328.0 |
| Glen Allan | 43.68 | 80.71 | 404.0 |
| Hamilton A | 43.17 | 79.93 | 238.0 |
| Ilderton | 43.05 | 81.43 | 266.7 |
| London A | 43.03 | 81.16 | 278.0 |
| Petrolia Town | 42.86 | 82.17 | 201.2 |
| Ridgetown | 42.45 | 81.88 | 210.3 |
| Sarnia | 43.00 | 82.32 | 191.0 |
| Stratford | 43.37 | 81.00 | 354.0 |
| St. Thomas | 42.78 | 81.21 | 209.0 |
| Tillsonburg | 42.86 | 80.72 | 270.0 |
| Waterloo Wellington | 43.46 | 80.38 | 317.0 |
| Woodstock | 43.14 | 80.77 | 282.0 |
| Wroxeter | 43.86 | 81.15 | 355.0 |

Table 2: Root mean square error values for selected climate indices from three weather generators at London A station

| Climate Index | SDSM | KnnCAD V4 | LARS-WG |
|---|---|---|---|
| Total monthly precipitation (mm) | 6.56 | 5.68 | 4.67 |
| Mean daily precipitation (mm) | 0.12 | 0.1 | 0.14 |
| Standard deviation daily precipitation (mm) | 0.44 | 0.4 | 0.32 |
| 95th percentile daily precipitation (mm) | 1.02 | 0.67 | 0.62 |
| 99th percentile daily precipitation (mm) | 4.06 | 2.85 | 2.09 |
| Mean wet spell length (days) | 0.29 | 0.15 | 0.49 |
| Mean dry spell length (days) | 0.54 | 0.38 | 0.39 |
| Max wet spell length (days) | 0.11 | 0.06 | 0.22 |
| Max dry spell length (days) | 0.23 | 0.16 | 0.38 |
| Mean maximum temperature (˚C) | 0.34 | 0.28 | 0.26 |
| Mean minimum temperature (˚C) | 0.31 | 0.25 | 0.3 |
| 99th percentile maximum temperature (˚C) | 1.44 | 0.46 | 1.89 |
| 95th percentile maximum temperature (˚C) | 0.78 | 0.4 | 2.17 |
| 5th percentile maximum temperature (˚C) | 0.61 | 0.44 | 1.46 |
| 1st percentile maximum temperature (˚C) | 1.35 | 0.42 | 1.53 |
| 99th percentile minimum temperature (˚C) | 1.49 | 0.58 | 2.33 |
| 95th percentile minimum temperature (˚C) | 0.82 | 0.3 | 1.58 |
| 5th percentile minimum temperature (˚C) | 0.44 | 0.33 | 1.44 |
| 1st percentile minimum temperature (˚C) | 1.14 | 0.22 | 1.16 |

Table 3: Historical and simulated mean values of CLIMDEX climate extreme indices from SDSM, KnnCAD V4 and LARS-WG.

| CLIMDEX Extreme Index | Description | Historical | SDSM | KnnCAD | LARSWG |
|---|---|---|---|---|---|
| PRCPTOT (mm) | Annual total precipitation | 1008.40 | 1004.28 | 973.54 | 1002.16 |
| CDD (days) | Maximum dry spell length (PPT<1mm) | 14.37 | 13.73 | 14.80 | 15.48 |
| CWD (days) | Maximum wet spell length (PPT>1mm) | 6.48 | 6.97 | 6.04 | 6.16 |
| R10mm (days) | Annual count of days with PPT<10mm | 32.00 | 30.88 | 30.46 | 31.23 |
| R20mm (days) | Annual count of days with PPT<20mm | 10.30 | 9.06 | 10.18 | 10.77 |
| R25mm (days) | Annual count of days with PPT<25mm | 6.44 | 5.33 | 6.24 | 6.79 |
| R95p (mm) | Annual total precipitation on days where PPT > 95th percentile historical | 231.20 | 189.99 | 229.17 | 250.91 |
| R99p (mm) | Annual total precipitation on days where PPT > 99th percentile historical | 67.74 | 51.55 | 71.73 | 86.09 |
| RX1day (mm) | Monthly maximum 1-day PPT | 49.71 | 51.03 | 55.61 | 56.09 |
| RX5day (mm) | Monthly maximum consecutive 5-day PPT | 79.39 | 79.71 | 84.29 | 82.44 |
| SDII (mm/time) | Simple precipitation intensity index | 7.85 | 7.48 | 7.76 | 8.00 |
| Tmax (c) | Monthly mean TMAX | 14.33 | 12.86 | 12.62 | 12.52 |
| Tmin (c) | Monthly mean TMIN | 4.81 | 2.76 | 2.90 | 2.82 |
| DTR (c) | Dinurial Temperature Range (Monthly mean difference between TMAX and TMIN) | 9.51 | 10.09 | 9.72 | 9.70 |
| FD (days) | Frost Days (Annual count of days with TMIN<0) | 112.56 | 143.15 | 143.09 | 146.20 |
| ID0 (days) | Icing Days (Annual count of days with TMAX<0) | 44.59 | 55.04 | 58.80 | 55.29 |
| SU (days) | Summer Days (Annual count of days with TMAX>25) | 87.00 | 61.40 | 62.96 | 55.88 |
| TR (days) | Tropical nights (TMIN>20) | 13.89 | 5.56 | 5.69 | 3.17 |
| WSDI (days) | Warm spell duration index (Annual count of 6-day periods with TMAX > 90th percentile) | 2.78 | 0.11 | 0.36 | 0.11 |
| CSDI (days) | Cold spell duration index (Annual count of 6 day periods with TMIN < 10th percentile) | 0.26 | 1.16 | 7.31 | 0.69 |

**Figures**

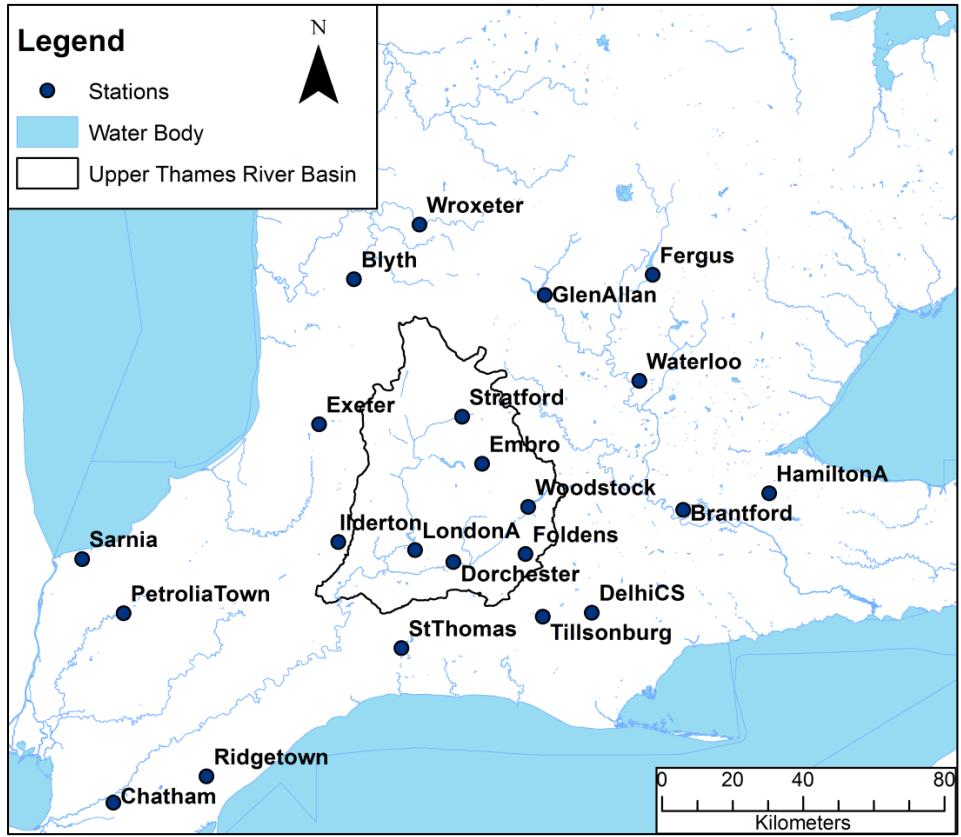

Figure 1: The Upper Thames River Basin

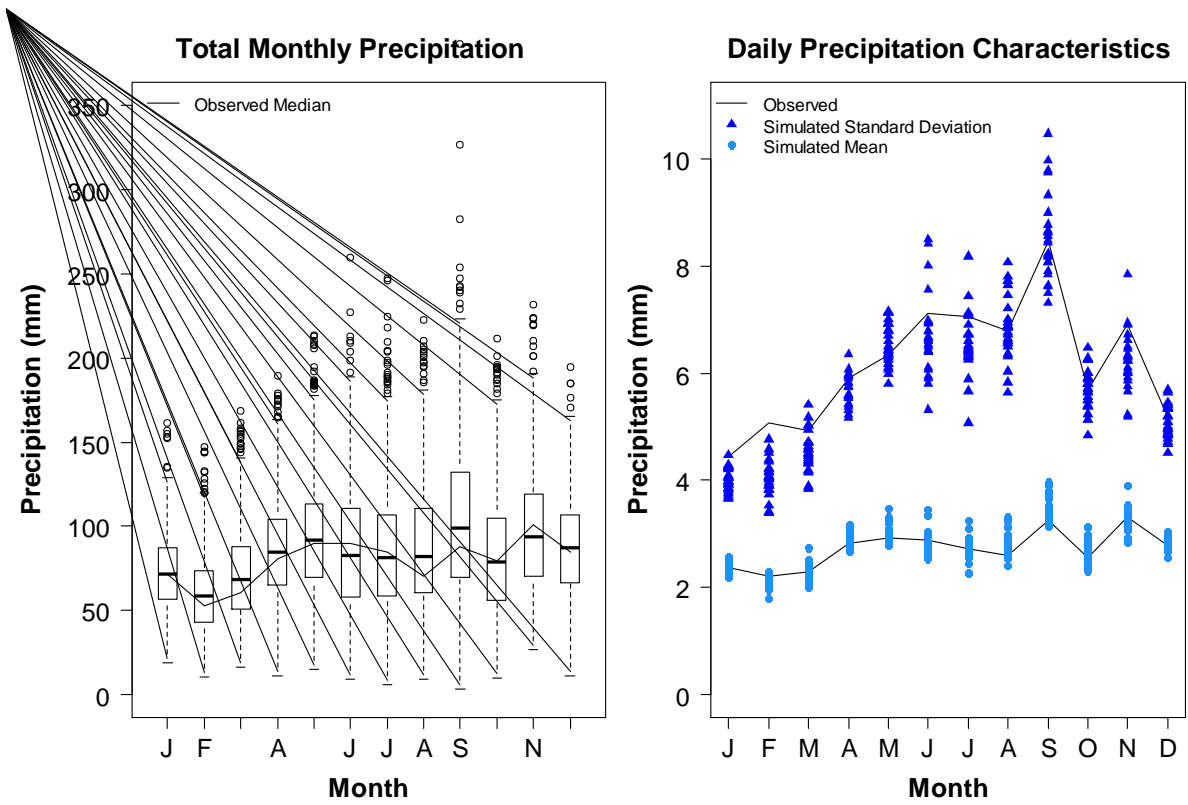

Figure 2: Simulated and observed total monthly precipitation and daily precipitation characteristics at London A

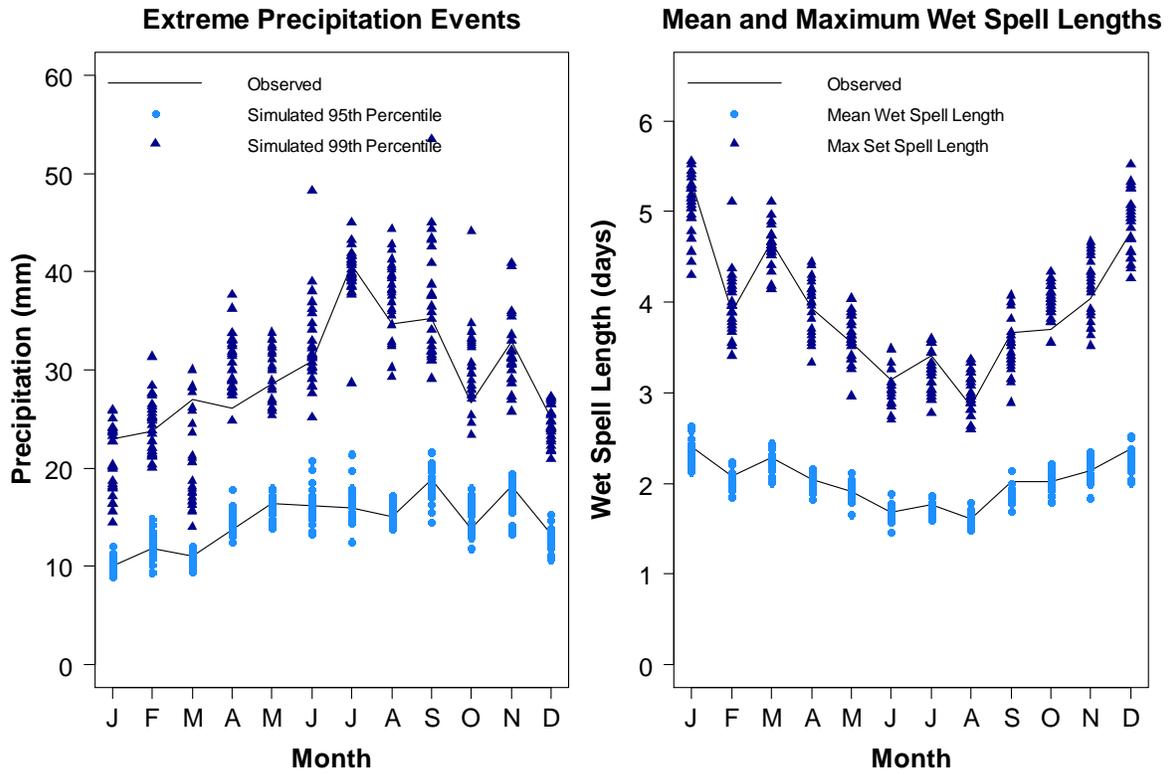

Figure 3: Simulated and observed extreme daily precipitation and wet spell lengths at London A

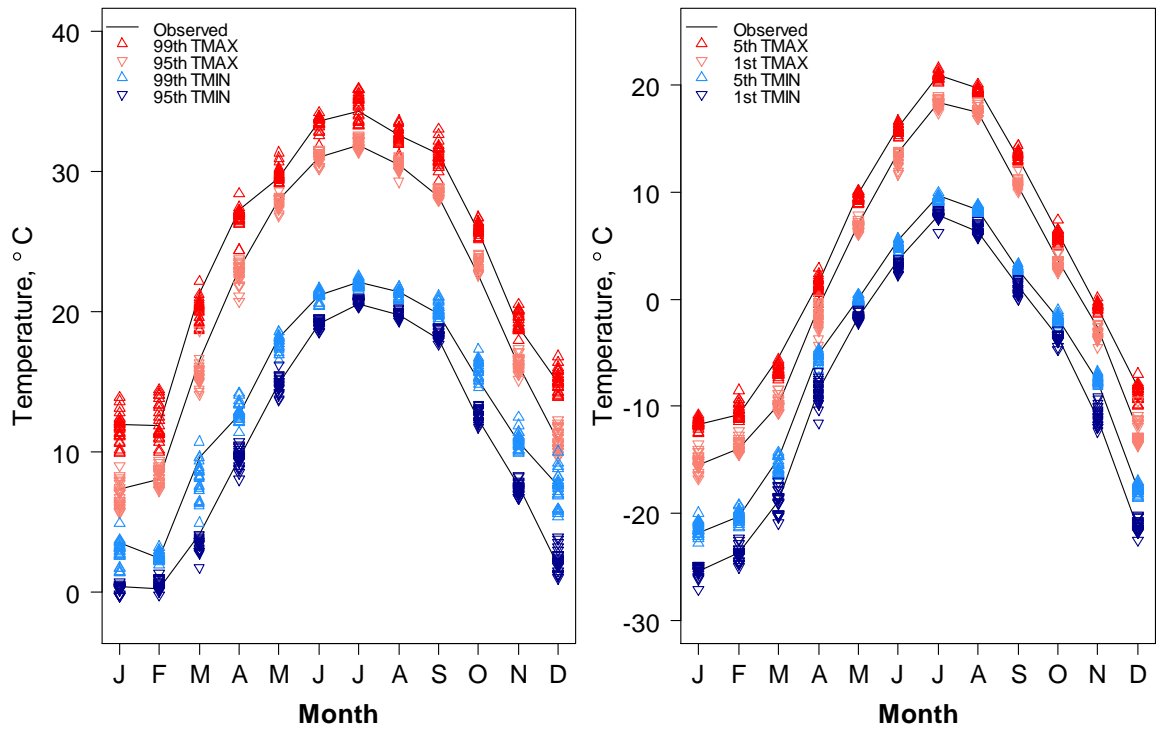

Figure 4: Simulated and observed extreme daily maximum and minimum temperatures at London A

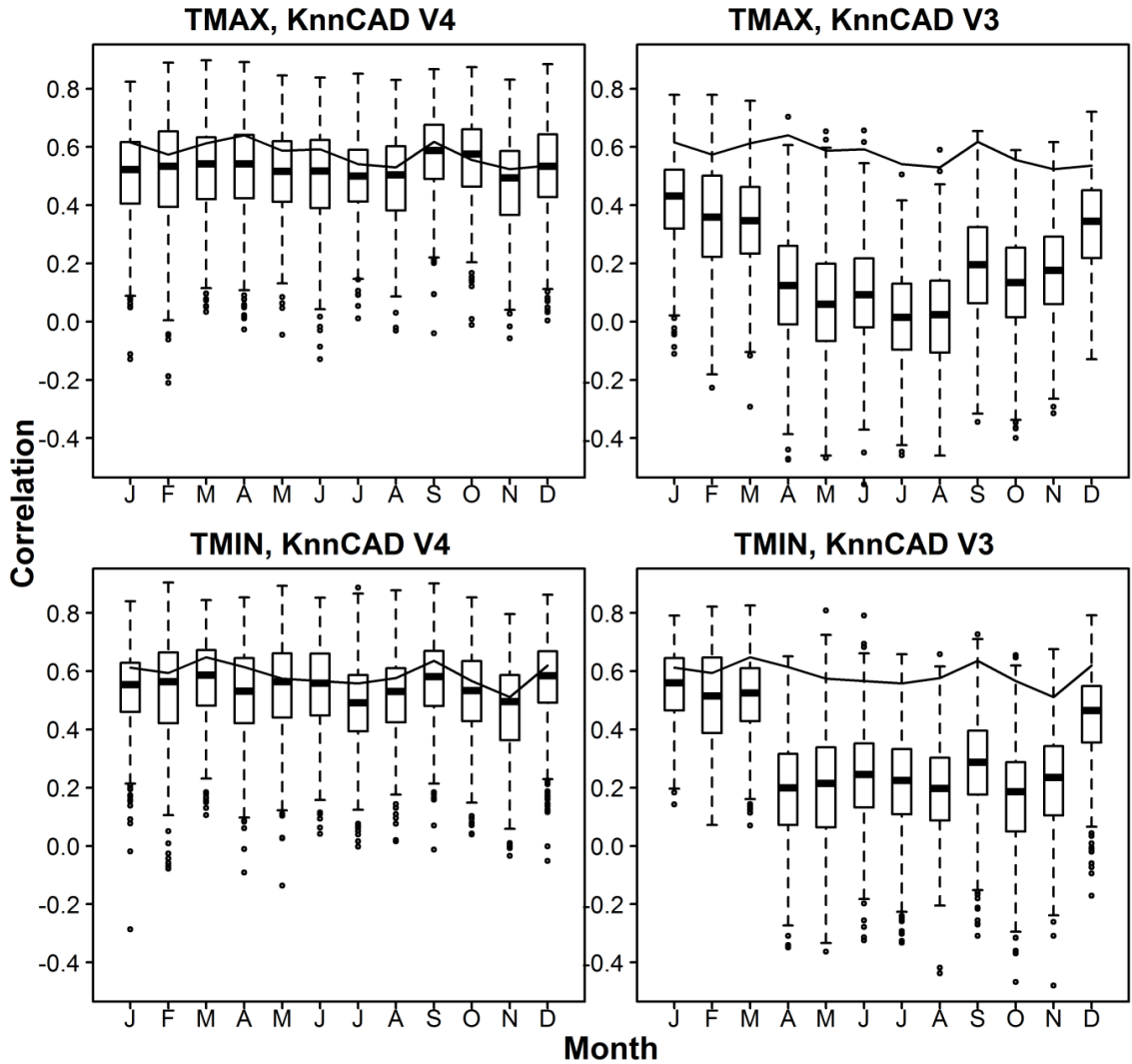

Figure 5: Simulated and observed lag-1 autocorrelations in maximum and minimum temperatures at London A from KnnCAD Versions 3 and 4.

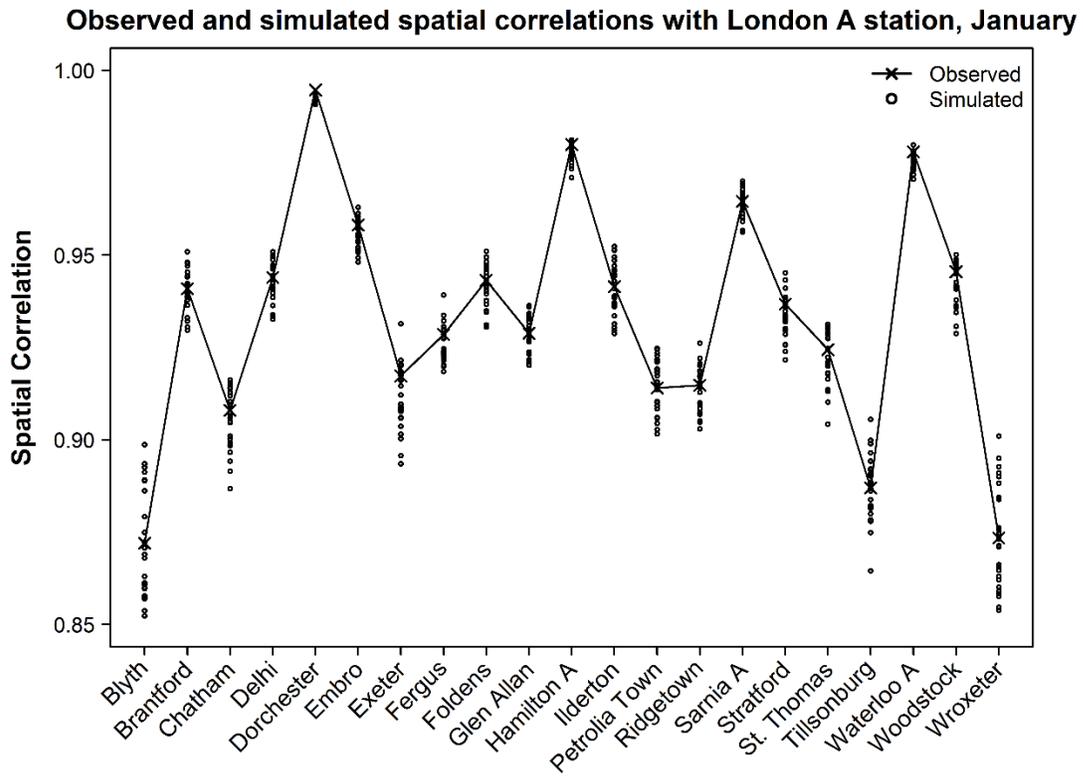

Figure 6: Spatial correlations of maximum temperatures for station pairs with London A for January.

**RX1d:** Monthly Maximum 1-Day Precipitationt Event

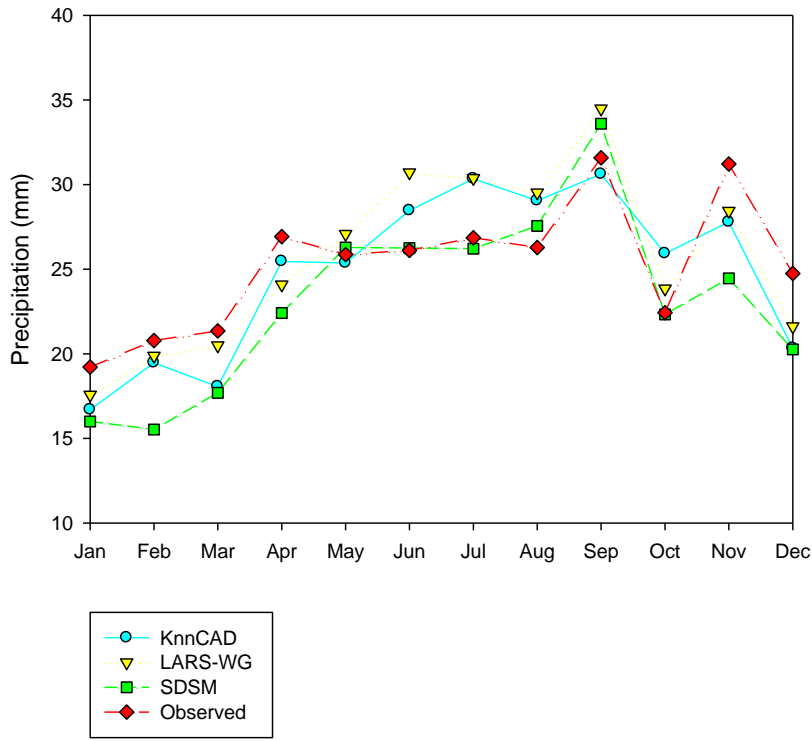

Figure 7: Monthly maximum 1-day precipitation amounts from the observed data and the simulated KnnCAD V4, LARS-WG and SDSM results.

**RX5d:** Monthly Maximum 5-Day Precipitation Event

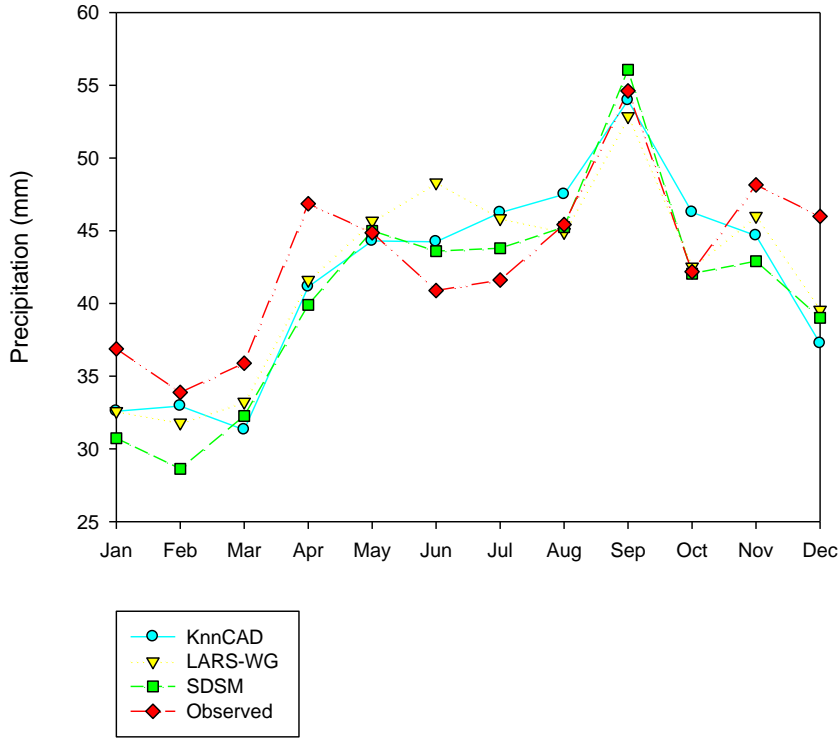

Figure 8: Monthly maximum 5-day consecutive precipitation amounts from the observed data and the simulated KnnCAD V4, LARS-WG and SDSM results.

**DTR:** Monthly Daily Temperature Range

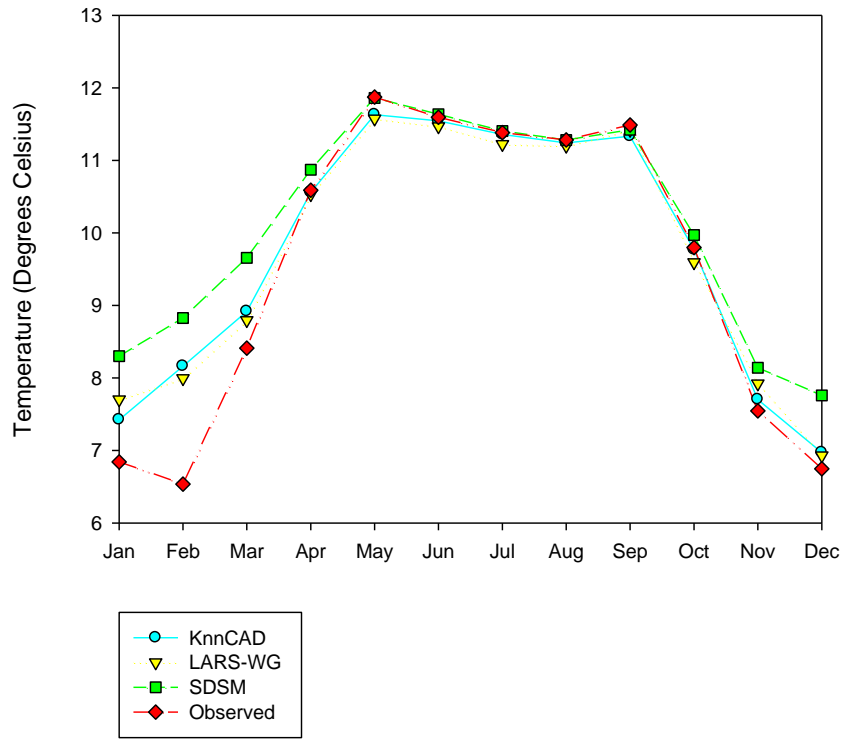

Figure 9: Monthly daily temperature range from the observed data and the simulated KnnCAD V4, LARS-WG and SDSM results.